\documentclass[twocolumn,amsmath,amssymb,prb,groupedaddress,superscriptaddress,reprint]{revtex4}
\usepackage{amsmath}
\usepackage{amssymb}
\usepackage{longtable}
\usepackage[dvips]{graphicx}
\usepackage{setspace}
\usepackage{color} 
\usepackage{moreverb}
\usepackage{subfigure}
\usepackage{bbm}

\newcommand{\InGaAs}[0]{$\text{In}_{0.5}$\hspace{0pt}$\text{Ga}_{0.5}$\hspace{0pt}$\text{As/}$\hspace{0pt}$\text{GaAs}$ }

\begin{document}
\title{Electric-field Manipulation of the Land\'e $g$ Tensor of Holes in
  \InGaAs Self-assembled Quantum Dots}
\author{Joseph Pingenot}
\affiliation{Center for Semiconductor Physics in Nanostructures, The
  University of Oklahoma, Norman, OK 73019}
\affiliation{Optical Science and Technology Center and Department of
  Physics and Astronomy, University of Iowa, Iowa City, IA 52242}
\author{Craig E. Pryor}
\affiliation{Optical Science and Technology Center and Department of
  Physics and Astronomy, University of Iowa, Iowa City, IA 52242}
\author{Michael E. Flatt\'e}
\affiliation{Optical Science and Technology Center and Department of
  Physics and Astronomy, University of Iowa, Iowa City, IA 52242}
\date{\today}
\begin{abstract}
The effect of an electric field on spin precession in \InGaAs self-assembled quantum dots is calculated using 
multiband real-space envelope-function theory. The dependence of the Land\'e $g$ tensor on electric fields should permit high-frequency $g$ tensor modulation resonance, as well as direct, nonresonant electric-field control of the hole spin. Subharmonic resonances have also been found in $g$ tensor modulation resonance of the holes, due to the strong quadratic dependence of components of the hole $g$ tensor on the electric field.
\end{abstract}
\maketitle

%Intro

\section{Introduction}
Since the initial proposals of spin-based quantum information processing\cite{Kane1998,Loss1998} in solids, methods of controlling individual spins in a scalable fashion have drawn considerable attention\cite{Awschalom2002,Hanson2007,Awschalom2007}. An especially attractive method for independently controlling spins spaced much closer than an optical spot size is to locally modify some of the spin's properties with an electric field. This approach's advantages include well-developed and commercial techniques for generating large arrays of independently controllable electric gates. Initial proposals focused on modifying the resonant frequency of a spin in a magnetic field using spatially-dependent $g$ factors\cite{Loss1998} or hyperfine fields\cite{Kane1998}, however this approach requires a spatially-extended always-on microwave field with which local spins are brought in and out of resonance.  A superior approach replaces electric-field control of a resonance frequency with electric-field control of the magnetic or pseudomagnetic field experienced by the spin, which can produce spin resonance when the electric field is oscillated at the resonance frequency.   Electric-field control of the magnetic field has been demonstrated by moving an electron back and forth (using electric field control of the location of quantum confinement of a quantum dot) in an inhomogeneous magnetic field\cite{Pioro2008}. Electric-field control of a pseudomagnetic field has largely focused on modifying the orbital moment of an electron, which then affects the $g$ tensor through the spin-orbit interaction; this approach goes by the name $g$ tensor modulation resonance, and has been demonstrated in quantum wells\cite{Kato2003} and also proposed for electron spin manipulation in quantum dots\cite{Pingenot2008}, for donor-bound electrons\cite{De2009}, and for holes in quantum dot molecules\cite{Andlauer2009}.

Advantages of using hole spins instead of electron spins for $g$ tensor modulation resonance include the larger spin-orbit interaction and orbital angular momenta in the valence band,  the stronger dependence (and asymmetry) of hole $g$ tensors on structural shape\cite{Nakaoka2004,Pryor2006,Kim2009}, and possibly reduced hyperfine interactions. The larger spin-orbit interaction for holes and larger orbital angular momenta lead to larger hole $g$ tensor components. If the fractional change in orbital angular momentum with applied electric field is comparable for electrons and holes, then the variation in $g$ tensor components should be much larger for holes, leading to much more rapid spin manipulation. To achieve rapid $g$ tensor modulation resonance it is especially advantageous to have one tensor component that changes as the other is larger unaffected\cite{Pingenot2008}. This can be achieved by considering a highly anisotropic dot shape. 
Zero in-plane $g$ tensor components were predicted for holes in self-assembled InAs/GaAs quantum dots with a circular footprint\cite{Pryor2006,Sheng2006}, although once the footprint becomes elliptical these $g$ tensor components grow rapidly\cite{Pryor2006}.   Other calculations have shown non-zero in-plane $g$ tensor components in
Ge/Si,\cite{Nenashev2003} and III-V nanowhisker\cite{De2007} quantum dots as well
as in a two-dot InAs/GaAs quantum dot molecule\cite{Andlauer2009}. Experimental measurements have seen small, but non-zero in-plane hole $g$ factors in InP/InGaP\cite{Yugova2002}, CdSe/ZnSe\cite{Koudinov2004} and InGaAs/GaAs quantum dots\cite{Yugova2007}. 
As an extreme example, in a Ge/Si nanowire quantum dot, electrical control of the  $g$ tensor component parallel to the wire has been experimentally realized by varying
the voltage on the electrostatic gates which define the
dot\cite{Roddaro2008}. Another extreme asymmetric structure is the vertically-coupled quantum-dot molecule, in which efficient hole $g$ tensor modulation by electric fields has been predicted\cite{Andlauer2009}, including the possibility of spin echo\cite{Roloff2010}.

Hole spins have very short lifetimes in bulk, due to angular-momentum mixing and state degeneracy typically found in the valence band\cite{Uenoyama1990}, but this lifetime
increases in quantum dots up to the same order of magnitude as electron spin lifetimes\cite{Gundogdu2005}.  Because the hole Bloch functions have
$p$-like character, the contact term of the hyperfine coupling to the nuclear spin
is zero, leaving only orbital hyperfine coupling to the nuclear
spin. The resulting hyperfine interaction has been shown to be highly anisotropic
for holes in quantum dots, with a pure heavy-hole state with
the pseudospin aligned along the $\hat z$ axis having little or no
hyperfine coupling\cite{Testelin2009}. Recent experiments have found hole $T_2^*$
times of $\sim 100-490$~ns\cite{Brunner2009}. In order to make a
quantum computer, at least $10^4$ operations must be performed during
the decoherence time.\cite{Preskill1998}  These recent experimental
$T_2$ values suggest a minimum spin manipulation time no shorter than 10-50~ps.

% Additionally, an nonresonant spin
%manipulation, which is capable of full Bloch-sphere manipulation, is
%possible when one or two (but not three) g-tensor components change
%sign.\cite{Pingenot2008}

Here we predict that both resonant and nonresonant spin manipulation is
possible in holes in a single \InGaAs quantum dot using a single
vertical electrical gate, and that the spin manipulation times are more rapid than those for electron spins in the same quantum dots. We also find that the nonlinearity of the $g$ tensor components with applied electric field is highly anisotropic, with a much stronger quadratic electric field dependence of the $g$ tensor component parallel to the electric field direction than perpendicular to it. Such quadratic electric field dependences have previously been studied for systems of high symmetry, like donor states\cite{De2009}, for which the linear electric field dependence vanishes. Here we find that these nonlinear $g$ tensor components generate highly anisotropic {\it subharmonic} resonances which effectively manipulate the hole spin. For magnetic fields of $\sim 5-10$~T the spin manipulation times are of the order of 20-30~ps for electric fields $\sim 150-200$~kV/cm.

\section{Hole $g$ tensor components in a single quantum dot}

\subsection{Theoretical method}

We have computed  $g$ tensors for holes confined in lens-shaped \InGaAs self-assembled quantum dots as a function of an electric field applied along the growth direction and ranging from $-150$~kV/cm to $150$~kV/cm.
The states were calculated using  an eight-band $\bf k\cdot p$ strain-dependent Hamiltonian\cite{Bahder1990} in the envelope approximation using finite differences on a real-space grid\cite{Pryor1998}. 
%The magnetic field was coupled to the envelope functions by phases on the differencing operators, and to the spin using a Pauli term\cite{Pryor2006}.  
This method has been used previously to calculate $g$ tensors of electrons in both quantum dots\cite{Pryor2006, Pingenot2008} and bound to donors\cite{De2009}.

The large electric fields considered here would ionize an electron in these dots\cite{Pingenot2008}, however holes remain confined due to their large effective mass.
%Another important difference between electrons and holes in a quantum dot is their spin. 
Whereas bulk holes have $J=3/2$, in a self-assembled dot the geometric asymmetry and strain break the four-fold degeneracy, resulting in doubly degenerate levels that are mixtures of heavy and light holes.
Although the lowest hole state is a doublet (mostly heavy hole, with small light hole components),  none of the eight  envelope functions  is identically zero and therefore the doublet energies will be split by a magnetic field pointing in any direction.
The hole $g$ tensor component for a magnetic field $B$  applied in a  direction $\bf \hat\alpha$ was
found by calculating the splitting  $\Delta E$ to obtain
$  \left| g_{\bf \hat \alpha} \right| = \Delta E/{3 \mu_B B}$,
where $\mu_B$ is the Bohr magneton.
We follow the convention for holes of including a factor of $3$ in the denominator (reflecting $m_J=\pm 3/2$) even though the states form a two-level system and are not pure heavy hole.
 The sign of  $g_{\bf \hat \alpha}$ was determined from the spin orientation of the ground state wave function, with $ g_{\bf \hat \alpha}  > 0 $ for the spin pointing antiparallel to $\bf B$.
Since  the lens-shaped dots were elongated along the $[110]$ direction the principal axes were $[110]$, $[1\overline{1}0]$, and $[001]$.
Spin splittings were computed for $\bf B$ along each of these symmetry directions to obtain each $g_{\bf \hat \alpha}$.

%The $g$ tensor component for a magnetic field applied along a direction $\bf \hat \alpha$ was found by computing the spin splitting of the lowest energy hole state, for which the $m_j = \pm\frac{3}{2}$ heavy hole was the largest component of the envelope function.
 
% The lens-shaped dots have mirror symmetry along their major and minor axes perpendicular to the growth direction, so the induced magnetic moment of the hole is parallel to the applied magnetic field when ${\bf B}$ is applied along the three symmetry directions: the growth direction ($[001]$), the major axis direction ($[1\overline{1}0]$)and the minor axis direction ($[110]$). Thus each $g$ tensor component was found by applying a magnetic field along the corresponding symmetry axis. 

\subsection{$g$ tensor dependence on electric field}

\begin{figure}[h!t]
  \begin{center}
      \includegraphics[width=\columnwidth]{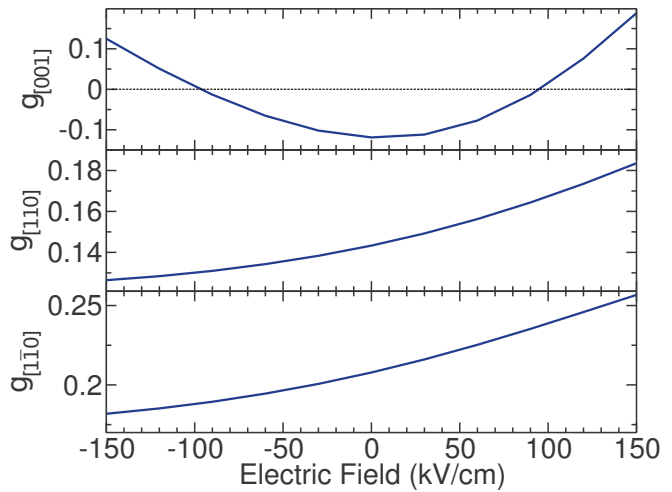}
      \caption{(color online)
        $g$ tensor of the uppermost valence state in an
        \InGaAs quantum dot as a function of
        electric field at 0K.  The dot has a height of 6.2nm, footprint length along
        the $[110]$ direction of 21.6nm, and footprint length along
        the $[1\overline{1}0]$ direction of 32.8nm.  
        Of particular  note is the sign change of the $g$ tensor component along the $[001]$ direction.
        \label{fig:g-tensor}
      }
  \end{center}
\end{figure}

Earlier investigations of electron spin manipulation using an electrically controlled $g$ tensor identified the importance of having a $g$ tensor component that changes sign with electric field\cite{Pingenot2008}.
If the sign of one $g_{\bf \hat \alpha}$ can be changed with an applied electric field, then it is possible to precess the spin to point in an arbitrary direction by only changing the electric field (in the presence of a suitably oriented static magnetic field).
 % Such a change in sign permits the spin precession vector to be rotated 90$^{\rm o}$ with an applied electric field. 
%The parameter space of greatest electric field control can be identified by investigating dots near the transition between positive and negative $g$ tensor component values due to changes in dot size or height. 
We have identified a dot geometry having the desired sign change, with a  height of $6.2~\rm nm$,  a $21.6~\rm nm $ base  along the minor axis ($[110]$ direction) and $32.8\rm ~nm$  along the major axis ($[1\overline{1}0]$ direction). 
 The $g$ tensor of this dot as a function of electric field is plotted in Fig.~\ref{fig:g-tensor}. 
The most interesting component is $g_{[001]}$ which is concave upward with a minimum value $g_{[001]} \approx -0.1$ at $E \approx 0$ and $g_{[001]} \approx 0$ at $E \approx \pm 100~ \rm kV/cm$.
%The $g$ tensor component along the $[001]$ direction in Fig.~\ref{fig:g-tensor} decreases as the electric field goes from positive (i.e. directed along $[001]$) to negative (i.e. directed along $[00\overline{1}]$), and then increases again and crosses $g_{[001]}=0$ as the electric field continues to increase. 
Therefore, there are two possible electric field values around which nonresonant spin manipulation may be done over the entire Bloch sphere as described in Ref.~\onlinecite{Pingenot2008}. 
This double sign change limits the range over which such  spin manipulation may be performed, as the limited range of $|g_{[001]}|$ constrains the effective spin precession frequency. 
However, the field ranges involved correspond roughly to the breakdown fields in GaAs, so larger electric fields are unlikely to be practical. 

In Fig.~\ref{fig:g-tensor} we see a significant nonlinearity in $g_{[001]}$, whereas the nonlinearities  in $g_{[110]}$ and $g_{[1\overline{1}0]}$ are much smaller.
% Although a small quadratic dependence on electric field can be seen in $g_{[110]}$ and $g_{[1\overline{1}0]}$, the dominant trend is a linear dependence on electric field. 
These nonlinearities were parameterized by fitting a second order polynomial in $E$ to the $g$ tensor components, resulting in the coefficients given in in Table~\ref{tab:coefficients}. 
\begingroup
\squeezetable 
\begin{table}[h!]
\begin{tabular}{||c||c|c|c||}
  \hline
  ${\bf \hat \alpha}$ & ${[001]}$ & ${[110]}$ & ${[1\overline{1}0]}$ \\
\hline
  \hline
  $c$& -0.115 & 0.143 & 0.208 \\
  \hline
  $b$  (cm/kV)& $1.15\times10^{-4}$ & $1.88\times10^{-4}$ & $2.52\times10^{-4}$ \\
  \hline
  $a$  $\rm (cm^2/kV^2)$& $1.23\times10^{-5}$ & $5.20\times10^{-7}$ &
  $5.08\times10^{-7}$ \\ 
  \hline
%  \hline
%  $b E$ & 0.0173 & 0.0282 & 0.0378 \\
%  \hline
%  $a E^2$ & 0.276 & 0.0117 & 0.0114 \\
%  \hline
%  \hline
\end{tabular}
\caption{
Coefficients of a fit of the $g$ tensor in
  Fig.~\ref{fig:g-tensor} to $g_{\bf \hat \alpha} = a_{\bf \hat \alpha}E^2 + b_{\bf \hat \alpha}E + c_{\bf \hat \alpha}$, where $E$
  is the electric field component along $[001]$. 
\label{tab:coefficients}
}
\end{table}
\endgroup

\section{Spin manipulation using $g$ tensor modulation with an electric field}

\subsection{Nonresonant hole spin manipulation with an electric field}
%Non-res manip
We first consider nonresonant spin precession using the technique developed in  Ref. \onlinecite{Pingenot2008}.
By applying two different  electric fields ($E_1$ and $E_2$) along the growth direction, precession around two orthogonal axes may be obtained. 
This approach requires the $g$ tensor component along one symmetry direction (here $[001]$) to change sign as a function of $E$. 
For a given $E_1$ and $E_2$ such that $g_{[001]}(E_1)~g_{[001]}(E_2)<0$, the magnetic field direction required  to obtain orthogonal spin precession axes $\bf \Omega = {\bf \tilde{g} }\cdot {\bf B}$ is determined by the condition
\begin{equation}
({\bf \tilde{g} }(E_1)\cdot {\bf  B})\cdot ({\bf \tilde{g} }(E_2)\cdot {\bf  B})=0.
\end{equation}
The optimal solution is determined by maximizing $|\Omega(E_1)|$ subject to  $|\Omega(E_1)| = |\Omega(E_2)|$ and $|E_{1,2}| < 150 ~\rm kV/cm$ (to avoid breakdown).
For the dot geometry corresponding to Fig.~\ref{fig:g-tensor}    we obtain $E_1 = -150\rm ~kV/cm$ and $E_2=3.1\rm ~kV/cm$.
%%%%   this is bad:
The optimal magnetic field angle ($0.24\pi$) is nearly $\pi/4$, measured from the [001] axis towards the [110] axis.
For a magnetic field of $5$~Tesla the time for the spin to precess by $\pi$  is 18~ps.

\subsection{Resonant hole spin manipulation with an electric field}
%Resonant manip
For a $g$ tensor that depends linearly on the electric field, resonances occur when the oscillation frequency of the electric field match the Larmor frequency of the spin in the static magnetic field.
However, when the $g$ tensor depends nonlinearly on the electric field, resonance may occur at subharmonics of the Larmor frequency\cite{De2009}. 
The strongly nonlinear behavior of the $g$ tensor for holes, evident in Fig.~\ref{fig:g-tensor} and parametrized in Table~\ref{tab:coefficients}, produces such subharmonic resonances.
For an applied field $E(t) = E_{dc}+E_{ac} \sin ( \omega t )$, the response amplitudes are
% sinusoidal applied electric field with amplitude $E_{AC}$ oscillating about $E_{DC}$,
%\begin{align*}
\begin{eqnarray}
  \Omega_{\bf \hat \alpha}(t) &=& B_{\bf \hat \alpha}\left(a_{\bf \hat \alpha} E^2(t) \sin^2(\omega t) +
  b_{\bf \hat \alpha} E(t) \sin(\omega t) + c_{\bf \hat \alpha}\right)\nonumber \\
  &=& \Omega_{0,\bf \hat \alpha} + \Omega_{1,\bf \hat \alpha} \sin(\omega t) - \Omega_{2,\bf \hat \alpha}  \cos(2 \omega t),
%\end{align*}
\end{eqnarray}
where 
\begin{align}
  \Omega_{0,\bf \hat \alpha} &= B_{\bf \hat \alpha}\left(\left[E_{dc}^2 + E_{ac}^2/2 \right]a_{\bf \hat \alpha} + E_{dc} b_{\bf \hat \alpha} + c_{\bf \hat \alpha}\right),\\
  \Omega_{1,\bf \hat \alpha} &= B_{\bf \hat \alpha}E_{ac}\left(2E_{dc} a_{\bf \hat \alpha} + b_{\bf \hat \alpha}\right),\\
  \Omega_{2,\bf \hat \alpha} &= B_{\bf \hat \alpha}E_{ac}^2a_{\bf \hat \alpha}/2.
\end{align}
$\Omega_1$ is the response at the fundamental $\omega=\omega_0$  and  $\Omega_2$ is the response at the subharmonic $\omega=\omega_0/2$, where $\omega_0$ is the Larmor precession frequency.
Higher order polynomial dependences ($e.g.$~cubic or quartic) of the $g$ tensor on the electric field
will result in resonances  at additional subharmonics of the Larmor frequency. Higher-order effects,
including the counter-rotating components of the oscillating
transverse component of the spin precession vector, will bring
additional shifts of the lower-order resonances and bring in
resonances at other multiples of the Larmor
frequency.\cite{Abragam1961}

The precession rates for the subharmonic and fundamental resonances
were calculated to first order for the full range of magnetic field angles,
and for electric field amplitudes less than the breakdown field $\sim 200$~kV/cm.
The  Rabi frequency associated with the fundamental resonance was found for an electric field
oscillating about $E_{dc} = 0$ as a function of applied magnetic field
direction, $\phi$, in the $[001]$-$[1\overline{1}0]$ plane and as a
function of electric field amplitude $E_{ac}$. For any given value
of $E$, an optimal  magnetic field direction was  found, corresponding to the largest Rabi frequency.
% In $\epsilon$, however, the trend across the examined range ($\epsilon$ between 0 and 600kV/cm, far beyond breakdown voltage) was simply that a larger $\epsilon$ yielded a faster spin manipulation time. 
As a function of $E$, the optimum magnetic field
direction increases monotonically. At $200$~kV/cm the largest Rabi frequency for a magnetic field of 10~Tesla is 18~GHz at a
optimal magnetic field angle of $1.2$~radians from the $[001]$ axis. The time required for the spin to precess by $\pi$ in this configuration is $\sim 28$~ps.

\begin{figure}[h!t]
  \begin{center}
    \includegraphics[width=\columnwidth]{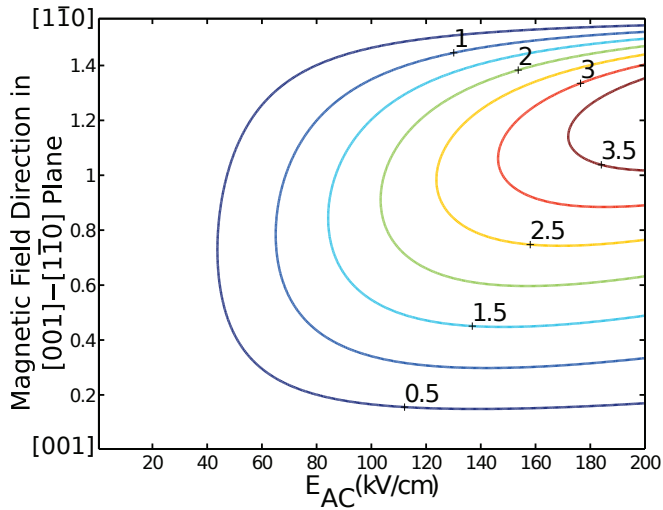}
      \caption{ (color online)
               Contour plot of Rabi frequency (in GHz) at quadratic resonance as a
        function of  AC electric field amplitude $E_{ac}$  and magnetic field angle, for $B = $1T.
        \label{fig:quadres}
      }
  \end{center}
\end{figure}

%Quadratic
The Rabi frequency for the subharmonic resonance was also found for an electric field
oscillating about $E_{dc} = 0$ as a function of applied magnetic field
direction $\phi$  and electric field amplitude $E$. As with the fundamental
resonance, a general trend of faster spin manipulation times at higher
$E$ was found through $E=200\text{kV/cm}$, peaking at an angle of 1.2 radians from the $[001]$ axis. 
At $B=10T$ the peak Rabi frequency was 39~GHz, corresponding to a minimal time for the spin to precess by $\pi$ of 13~ps.

%Conclusion
\section{Concluding Remarks}
We have examined theoretically the hole
$g$ tensors in \InGaAs quantum dots for possible application to hole
spin manipulation. 
%All InAs/GaAs quantum dot structures examined had the same $g$ tensor component signs. 
A structure was proposed for which one $g$ tensor component sign changed as a function of electric field
applied along the $[001]$ growth direction. The nonlinear $g$ tensor
dependence on applied electric field causes a subharmonic
resonance to appear at $\omega = \omega_0/2$, with higher-order
dependencies generating further subharmonics. The $g$ tensor for this
structure was used to calculate resonant and nonresonant spin
manipulation frequencies. For a magnetic field of 10~Tesla, the resonant spin manipulation method had an
optimal spin manipulation time (precession by $\pi$) of 28~ps at the fundamental resonance, or
13~ps at the subharmonic resonance. The nonresonant spin manipulation
frequency for a $5$~Tesla field was 18~ps.  Using the experimental values of $T_2$ from
\cite{Brunner2009}, approximately $10^4$ operations would be possible during a $T_2$ time.


\begin{thebibliography}{29}
\expandafter\ifx\csname natexlab\endcsname\relax\def\natexlab#1{#1}\fi
\expandafter\ifx\csname bibnamefont\endcsname\relax
  \def\bibnamefont#1{#1}\fi
\expandafter\ifx\csname bibfnamefont\endcsname\relax
  \def\bibfnamefont#1{#1}\fi
\expandafter\ifx\csname citenamefont\endcsname\relax
  \def\citenamefont#1{#1}\fi
\expandafter\ifx\csname url\endcsname\relax
  \def\url#1{\texttt{#1}}\fi
\expandafter\ifx\csname urlprefix\endcsname\relax\def\urlprefix{URL }\fi
\providecommand{\bibinfo}[2]{#2}
\providecommand{\eprint}[2][]{\url{#2}}

\bibitem[{\citenamefont{Kane}(1998)}]{Kane1998}
\bibinfo{author}{\bibfnamefont{B.~E.} \bibnamefont{Kane}},
  \bibinfo{journal}{Nature} \textbf{\bibinfo{volume}{393}},
  \bibinfo{pages}{133} (\bibinfo{year}{1998}).

\bibitem[{\citenamefont{Loss and DiVincenzo}(1998)}]{Loss1998}
\bibinfo{author}{\bibfnamefont{D.}~\bibnamefont{Loss}} \bibnamefont{and}
  \bibinfo{author}{\bibfnamefont{D.~P.} \bibnamefont{DiVincenzo}},
  \bibinfo{journal}{\pra} \textbf{\bibinfo{volume}{57}}, \bibinfo{pages}{120}
  (\bibinfo{year}{1998}).

\bibitem[{\citenamefont{Awschalom et~al.}(2002)\citenamefont{Awschalom,
  Samarth, and Loss}}]{Awschalom2002}
\bibinfo{editor}{\bibfnamefont{D.~D.} \bibnamefont{Awschalom}},
  \bibinfo{editor}{\bibfnamefont{N.}~\bibnamefont{Samarth}}, \bibnamefont{and}
  \bibinfo{editor}{\bibfnamefont{D.}~\bibnamefont{Loss}}, eds.,
  \emph{\bibinfo{title}{Semiconductor Spintronics and Quantum Computation}}
  (\bibinfo{publisher}{Springer Verlag}, \bibinfo{address}{Heidelberg},
  \bibinfo{year}{2002}).

\bibitem[{\citenamefont{Hanson et~al.}(2007)\citenamefont{Hanson, Kouwenhoven,
  Petta, Tarucha, and Vandersypen}}]{Hanson2007}
\bibinfo{author}{\bibfnamefont{R.}~\bibnamefont{Hanson}},
  \bibinfo{author}{\bibfnamefont{L.~P.} \bibnamefont{Kouwenhoven}},
  \bibinfo{author}{\bibfnamefont{J.~R.} \bibnamefont{Petta}},
  \bibinfo{author}{\bibfnamefont{S.}~\bibnamefont{Tarucha}}, \bibnamefont{and}
  \bibinfo{author}{\bibfnamefont{L.~M.~K.} \bibnamefont{Vandersypen}},
  \bibinfo{journal}{Reviews of Modern Physics} \textbf{\bibinfo{volume}{79}},
  \bibinfo{eid}{1217} (\bibinfo{year}{2007}).

\bibitem[{\citenamefont{Awschalom and Flatt\'e}(2007)}]{Awschalom2007}
\bibinfo{author}{\bibfnamefont{D.~D.} \bibnamefont{Awschalom}}
  \bibnamefont{and} \bibinfo{author}{\bibfnamefont{M.~E.}
  \bibnamefont{Flatt\'e}}, \bibinfo{journal}{Nature Physics}
  \textbf{\bibinfo{volume}{3}}, \bibinfo{pages}{153} (\bibinfo{year}{2007}).

\bibitem[{\citenamefont{Pioro-Ladri\`ere
  et~al.}(2008)\citenamefont{Pioro-Ladri\`ere, Obata, Tokura, Shin, Kubo,
  Yoshida, Taniyama, and Tarucha}}]{Pioro2008}
\bibinfo{author}{\bibfnamefont{M.}~\bibnamefont{Pioro-Ladri\`ere}},
  \bibinfo{author}{\bibfnamefont{T.}~\bibnamefont{Obata}},
  \bibinfo{author}{\bibfnamefont{Y.}~\bibnamefont{Tokura}},
  \bibinfo{author}{\bibfnamefont{Y.-S.} \bibnamefont{Shin}},
  \bibinfo{author}{\bibfnamefont{T.}~\bibnamefont{Kubo}},
  \bibinfo{author}{\bibfnamefont{K.}~\bibnamefont{Yoshida}},
  \bibinfo{author}{\bibfnamefont{T.}~\bibnamefont{Taniyama}}, \bibnamefont{and}
  \bibinfo{author}{\bibfnamefont{S.}~\bibnamefont{Tarucha}},
  \bibinfo{journal}{Nature Physics} \textbf{\bibinfo{volume}{4}},
  \bibinfo{pages}{776} (\bibinfo{year}{2008}).

\bibitem[{\citenamefont{Kato et~al.}(2003)\citenamefont{Kato, Myers, Gossard,
  Levy, and Awschalom}}]{Kato2003}
\bibinfo{author}{\bibfnamefont{Y.}~\bibnamefont{Kato}},
  \bibinfo{author}{\bibfnamefont{R.~C.} \bibnamefont{Myers}},
  \bibinfo{author}{\bibfnamefont{A.~C.} \bibnamefont{Gossard}},
  \bibinfo{author}{\bibfnamefont{J.}~\bibnamefont{Levy}}, \bibnamefont{and}
  \bibinfo{author}{\bibfnamefont{D.~D.} \bibnamefont{Awschalom}},
  \bibinfo{journal}{Science} \textbf{\bibinfo{volume}{299}},
  \bibinfo{pages}{1201} (\bibinfo{year}{2003}).

\bibitem[{\citenamefont{Pingenot et~al.}(2008)\citenamefont{Pingenot, Pryor,
  and Flatt\'{e}}}]{Pingenot2008}
\bibinfo{author}{\bibfnamefont{J.}~\bibnamefont{Pingenot}},
  \bibinfo{author}{\bibfnamefont{C.~E.} \bibnamefont{Pryor}}, \bibnamefont{and}
  \bibinfo{author}{\bibfnamefont{M.~E.} \bibnamefont{Flatt\'{e}}},
  \bibinfo{journal}{Appl. Phys. Lett.} \textbf{\bibinfo{volume}{92}},
  \bibinfo{eid}{222502} (\bibinfo{year}{2008}).

\bibitem[{\citenamefont{De et~al.}(2009)\citenamefont{De, Pryor, and
  Flatt\'{e}}}]{De2009}
\bibinfo{author}{\bibfnamefont{A.}~\bibnamefont{De}},
  \bibinfo{author}{\bibfnamefont{C.~E.} \bibnamefont{Pryor}}, \bibnamefont{and}
  \bibinfo{author}{\bibfnamefont{M.~E.} \bibnamefont{Flatt\'{e}}},
  \bibinfo{journal}{Physical Review Letters} \textbf{\bibinfo{volume}{102}},
  \bibinfo{eid}{017603} (\bibinfo{year}{2009}).

\bibitem[{\citenamefont{Andlauer and Vogl}(2009)}]{Andlauer2009}
\bibinfo{author}{\bibfnamefont{T.}~\bibnamefont{Andlauer}} \bibnamefont{and}
  \bibinfo{author}{\bibfnamefont{P.}~\bibnamefont{Vogl}},
  \bibinfo{journal}{Phys. Rev. B} \textbf{\bibinfo{volume}{79}},
  \bibinfo{pages}{045307} (\bibinfo{year}{2009}).

\bibitem[{\citenamefont{Nakaoka et~al.}(2004)\citenamefont{Nakaoka, Saito,
  Tatebayashi, and Arakawa}}]{Nakaoka2004}
\bibinfo{author}{\bibfnamefont{T.}~\bibnamefont{Nakaoka}},
  \bibinfo{author}{\bibfnamefont{T.}~\bibnamefont{Saito}},
  \bibinfo{author}{\bibfnamefont{J.}~\bibnamefont{Tatebayashi}},
  \bibnamefont{and} \bibinfo{author}{\bibfnamefont{Y.}~\bibnamefont{Arakawa}},
  \bibinfo{journal}{Phys. Rev. B} \textbf{\bibinfo{volume}{70}},
  \bibinfo{pages}{235337} (\bibinfo{year}{2004}).

\bibitem[{\citenamefont{Pryor and Flatt\'e}(2006)}]{Pryor2006}
\bibinfo{author}{\bibfnamefont{C.~E.} \bibnamefont{Pryor}} \bibnamefont{and}
  \bibinfo{author}{\bibfnamefont{M.~E.} \bibnamefont{Flatt\'e}},
  \bibinfo{journal}{\apl} \textbf{\bibinfo{volume}{88}},
  \bibinfo{pages}{233108} (\bibinfo{year}{2006}).

\bibitem[{\citenamefont{Kim et~al.}(2009)\citenamefont{Kim, Sheng, Poole,
  Dalacu, Lefebvre, Lapointe, Reimer, Aers, and Williams}}]{Kim2009}
\bibinfo{author}{\bibfnamefont{D.}~\bibnamefont{Kim}},
  \bibinfo{author}{\bibfnamefont{W.}~\bibnamefont{Sheng}},
  \bibinfo{author}{\bibfnamefont{P.~J.} \bibnamefont{Poole}},
  \bibinfo{author}{\bibfnamefont{D.}~\bibnamefont{Dalacu}},
  \bibinfo{author}{\bibfnamefont{J.}~\bibnamefont{Lefebvre}},
  \bibinfo{author}{\bibfnamefont{J.}~\bibnamefont{Lapointe}},
  \bibinfo{author}{\bibfnamefont{M.~E.} \bibnamefont{Reimer}},
  \bibinfo{author}{\bibfnamefont{G.~C.} \bibnamefont{Aers}}, \bibnamefont{and}
  \bibinfo{author}{\bibfnamefont{R.~L.} \bibnamefont{Williams}},
  \bibinfo{journal}{Physical Review B} \textbf{\bibinfo{volume}{79}},
  \bibinfo{pages}{045310} (\bibinfo{year}{2009}).

\bibitem[{\citenamefont{Sheng and Hawrylak}(2006)}]{Sheng2006}
\bibinfo{author}{\bibfnamefont{W.}~\bibnamefont{Sheng}} \bibnamefont{and}
  \bibinfo{author}{\bibfnamefont{P.}~\bibnamefont{Hawrylak}},
  \bibinfo{journal}{Phys. Rev. B} \textbf{\bibinfo{volume}{73}},
  \bibinfo{pages}{125331} (\bibinfo{year}{2006}).

\bibitem[{\citenamefont{Nenashev et~al.}(2003)\citenamefont{Nenashev,
  Dvurechenskii, and Zinovieva}}]{Nenashev2003}
\bibinfo{author}{\bibfnamefont{A.~V.} \bibnamefont{Nenashev}},
  \bibinfo{author}{\bibfnamefont{A.~V.} \bibnamefont{Dvurechenskii}},
  \bibnamefont{and} \bibinfo{author}{\bibfnamefont{A.~F.}
  \bibnamefont{Zinovieva}}, \bibinfo{journal}{Phys. Rev. B}
  \textbf{\bibinfo{volume}{67}}, \bibinfo{pages}{205301}
  (\bibinfo{year}{2003}).

\bibitem[{\citenamefont{De and Pryor}(2007)}]{De2007}
\bibinfo{author}{\bibfnamefont{A.}~\bibnamefont{De}} \bibnamefont{and}
  \bibinfo{author}{\bibfnamefont{C.~E.} \bibnamefont{Pryor}},
  \bibinfo{journal}{Phys. Rev. B} \textbf{\bibinfo{volume}{76}},
  \bibinfo{pages}{155321} (\bibinfo{year}{2007}).

\bibitem[{\citenamefont{Yugova et~al.}(2002)\citenamefont{Yugova, Gerlovin,
  Davydov, Ignatiev, Kozin, Ren, Sugisaki, Sugou, and Masumoto}}]{Yugova2002}
\bibinfo{author}{\bibfnamefont{I.~A.} \bibnamefont{Yugova}},
  \bibinfo{author}{\bibfnamefont{I.~Y.} \bibnamefont{Gerlovin}},
  \bibinfo{author}{\bibfnamefont{V.~G.} \bibnamefont{Davydov}},
  \bibinfo{author}{\bibfnamefont{I.~V.} \bibnamefont{Ignatiev}},
  \bibinfo{author}{\bibfnamefont{I.~E.} \bibnamefont{Kozin}},
  \bibinfo{author}{\bibfnamefont{H.~W.} \bibnamefont{Ren}},
  \bibinfo{author}{\bibfnamefont{M.}~\bibnamefont{Sugisaki}},
  \bibinfo{author}{\bibfnamefont{S.}~\bibnamefont{Sugou}}, \bibnamefont{and}
  \bibinfo{author}{\bibfnamefont{Y.}~\bibnamefont{Masumoto}},
  \bibinfo{journal}{Phys. Rev. B} \textbf{\bibinfo{volume}{66}},
  \bibinfo{pages}{235312} (\bibinfo{year}{2002}).

\bibitem[{\citenamefont{Koudinov et~al.}(2004)\citenamefont{Koudinov, Akimov,
  Kusrayev, and Henneberger}}]{Koudinov2004}
\bibinfo{author}{\bibfnamefont{A.~V.} \bibnamefont{Koudinov}},
  \bibinfo{author}{\bibfnamefont{I.~A.} \bibnamefont{Akimov}},
  \bibinfo{author}{\bibfnamefont{Y.~G.} \bibnamefont{Kusrayev}},
  \bibnamefont{and}
  \bibinfo{author}{\bibfnamefont{F.}~\bibnamefont{Henneberger}},
  \bibinfo{journal}{Phys. Rev. B} \textbf{\bibinfo{volume}{70}},
  \bibinfo{pages}{241305} (\bibinfo{year}{2004}).

\bibitem[{\citenamefont{Yugova et~al.}(2007)\citenamefont{Yugova, Greilich,
  Zhukov, Yakovlev, Bayer, Reuter, and Wieck}}]{Yugova2007}
\bibinfo{author}{\bibfnamefont{I.~A.} \bibnamefont{Yugova}},
  \bibinfo{author}{\bibfnamefont{A.}~\bibnamefont{Greilich}},
  \bibinfo{author}{\bibfnamefont{E.~A.} \bibnamefont{Zhukov}},
  \bibinfo{author}{\bibfnamefont{D.~R.} \bibnamefont{Yakovlev}},
  \bibinfo{author}{\bibfnamefont{M.}~\bibnamefont{Bayer}},
  \bibinfo{author}{\bibfnamefont{D.}~\bibnamefont{Reuter}}, \bibnamefont{and}
  \bibinfo{author}{\bibfnamefont{A.~D.} \bibnamefont{Wieck}},
  \bibinfo{journal}{Phys. Rev. B} \textbf{\bibinfo{volume}{75}},
  \bibinfo{pages}{195325} (\bibinfo{year}{2007}).

\bibitem[{\citenamefont{Roddaro et~al.}(2008)\citenamefont{Roddaro, Fuhrer,
  Brusheim, Fasth, Xu, Samuelson, Xiang, and Lieber}}]{Roddaro2008}
\bibinfo{author}{\bibfnamefont{S.}~\bibnamefont{Roddaro}},
  \bibinfo{author}{\bibfnamefont{A.}~\bibnamefont{Fuhrer}},
  \bibinfo{author}{\bibfnamefont{P.}~\bibnamefont{Brusheim}},
  \bibinfo{author}{\bibfnamefont{C.}~\bibnamefont{Fasth}},
  \bibinfo{author}{\bibfnamefont{H.~Q.} \bibnamefont{Xu}},
  \bibinfo{author}{\bibfnamefont{L.}~\bibnamefont{Samuelson}},
  \bibinfo{author}{\bibfnamefont{J.}~\bibnamefont{Xiang}}, \bibnamefont{and}
  \bibinfo{author}{\bibfnamefont{C.~M.} \bibnamefont{Lieber}},
  \bibinfo{journal}{Physical Review Letters} \textbf{\bibinfo{volume}{101}},
  \bibinfo{pages}{186802} (\bibinfo{year}{2008}).

\bibitem[{\citenamefont{Roloff et~al.}(2010)\citenamefont{Roloff, P\"otz,
  Eissfeller, and Vogl}}]{Roloff2010}
\bibinfo{author}{\bibfnamefont{R.}~\bibnamefont{Roloff}},
  \bibinfo{author}{\bibfnamefont{W.}~\bibnamefont{P\"otz}},
  \bibinfo{author}{\bibfnamefont{T.}~\bibnamefont{Eissfeller}},
  \bibnamefont{and} \bibinfo{author}{\bibfnamefont{P.}~\bibnamefont{Vogl}}
  (\bibinfo{year}{2010}), \eprint{arxiv:1003.0897v1}.

\bibitem[{\citenamefont{Uenoyama and Sham}(1990)}]{Uenoyama1990}
\bibinfo{author}{\bibfnamefont{T.}~\bibnamefont{Uenoyama}} \bibnamefont{and}
  \bibinfo{author}{\bibfnamefont{L.~J.} \bibnamefont{Sham}},
  \bibinfo{journal}{Phys. Rev. Lett.} \textbf{\bibinfo{volume}{64}},
  \bibinfo{pages}{3070} (\bibinfo{year}{1990}).

\bibitem[{\citenamefont{G\"undo\u{g}du
  et~al.}(2005)\citenamefont{G\"undo\u{g}du, Hall, Koerperick, Pryor, Flatt\'e,
  Boggess, Shchekin, and Deppe}}]{Gundogdu2005}
\bibinfo{author}{\bibfnamefont{K.}~\bibnamefont{G\"undo\u{g}du}},
  \bibinfo{author}{\bibfnamefont{K.~C.} \bibnamefont{Hall}},
  \bibinfo{author}{\bibfnamefont{E.~J.} \bibnamefont{Koerperick}},
  \bibinfo{author}{\bibfnamefont{C.~E.} \bibnamefont{Pryor}},
  \bibinfo{author}{\bibfnamefont{M.~E.} \bibnamefont{Flatt\'e}},
  \bibinfo{author}{\bibfnamefont{T.~F.} \bibnamefont{Boggess}},
  \bibinfo{author}{\bibfnamefont{O.~B.} \bibnamefont{Shchekin}},
  \bibnamefont{and} \bibinfo{author}{\bibfnamefont{D.~G.} \bibnamefont{Deppe}},
  \bibinfo{journal}{Appl. Phys. Lett.} \textbf{\bibinfo{volume}{86}},
  \bibinfo{pages}{113111} (\bibinfo{year}{2005}).

\bibitem[{\citenamefont{Testelin et~al.}(2009)\citenamefont{Testelin, Bernadot,
  Eble, and Chamarro}}]{Testelin2009}
\bibinfo{author}{\bibfnamefont{C.}~\bibnamefont{Testelin}},
  \bibinfo{author}{\bibfnamefont{F.}~\bibnamefont{Bernadot}},
  \bibinfo{author}{\bibfnamefont{B.}~\bibnamefont{Eble}}, \bibnamefont{and}
  \bibinfo{author}{\bibfnamefont{M.}~\bibnamefont{Chamarro}},
  \bibinfo{journal}{Physical Review B} \textbf{\bibinfo{volume}{79}},
  \bibinfo{pages}{195440} (\bibinfo{year}{2009}).

\bibitem[{\citenamefont{Brunner et~al.}(2009)\citenamefont{Brunner, Geradot,
  Dalgarno, W\"ust, Karrai, Stolta, Petroff, and Warburton}}]{Brunner2009}
\bibinfo{author}{\bibfnamefont{D.}~\bibnamefont{Brunner}},
  \bibinfo{author}{\bibfnamefont{B.~D.} \bibnamefont{Geradot}},
  \bibinfo{author}{\bibfnamefont{P.~A.} \bibnamefont{Dalgarno}},
  \bibinfo{author}{\bibfnamefont{G.}~\bibnamefont{W\"ust}},
  \bibinfo{author}{\bibfnamefont{K.}~\bibnamefont{Karrai}},
  \bibinfo{author}{\bibfnamefont{N.~G.} \bibnamefont{Stolta}},
  \bibinfo{author}{\bibfnamefont{P.~M.} \bibnamefont{Petroff}},
  \bibnamefont{and} \bibinfo{author}{\bibfnamefont{R.~J.}
  \bibnamefont{Warburton}}, \bibinfo{journal}{Science}
  \textbf{\bibinfo{volume}{325}}, \bibinfo{pages}{70} (\bibinfo{year}{2009}).

\bibitem[{\citenamefont{Preskill}(1998)}]{Preskill1998}
\bibinfo{author}{\bibfnamefont{J.}~\bibnamefont{Preskill}}, in
  \emph{\bibinfo{booktitle}{Introduction to quantum computation and
  information}}, edited by \bibinfo{editor}{\bibfnamefont{H.-K.}
  \bibnamefont{Lo}}, \bibinfo{editor}{\bibfnamefont{S.}~\bibnamefont{Popescu}},
  \bibnamefont{and} \bibinfo{editor}{\bibfnamefont{T.}~\bibnamefont{Spiller}}
  (\bibinfo{publisher}{World Scientific}, \bibinfo{address}{Singapore},
  \bibinfo{year}{1998}), pp. \bibinfo{pages}{213--269}.

\bibitem[{\citenamefont{Bahder}(1990)}]{Bahder1990}
\bibinfo{author}{\bibfnamefont{T.~B.} \bibnamefont{Bahder}},
  \bibinfo{journal}{Phys. Rev. B} \textbf{\bibinfo{volume}{41}},
  \bibinfo{pages}{11992} (\bibinfo{year}{1990}).

\bibitem[{\citenamefont{Pryor}(1998)}]{Pryor1998}
\bibinfo{author}{\bibfnamefont{C.}~\bibnamefont{Pryor}},
  \bibinfo{journal}{Phys. Rev. B} \textbf{\bibinfo{volume}{57}},
  \bibinfo{pages}{7190} (\bibinfo{year}{1998}).

\bibitem[{\citenamefont{Abragam}(1961)}]{Abragam1961}
\bibinfo{author}{\bibfnamefont{A.}~\bibnamefont{Abragam}},
  \emph{\bibinfo{title}{Principles of Nuclear Magnetism}}
  (\bibinfo{publisher}{Oxford Science Publications}, \bibinfo{year}{1961}),
  ISBN \bibinfo{isbn}{019852014X}.

\end{thebibliography}
\end{document}